\documentclass{article}
\usepackage{amsmath,amssymb,mathtools,cite,graphicx,fancyhdr}
\usepackage{bbold,color,bm,caption}
\graphicspath{{Graphs/}}

\begin{document}
\lhead{Dialogues about geometry and light}
\rhead{D. Bermudez, J. Drori and U. Leonhardt}

\title{Dialogues about geometry and light}
\author{David Bermudez, Jonathan Drori and Ulf Leonhardt \\
{\sl Department of Physics of Complex Systems}\\
{\sl Weizmann Institute of Science, Rehovot 76100, Israel}
}

\date{}

\maketitle

\begin{abstract}
Throughout human history, people have used sight to learn about the world, but only in relatively recent times the science of light has been developed. Egyptians and Mesopotamians made the first known lenses out of quartz, giving birth to what was later known as {\it optics}. On the other hand, {\it geometry} is a branch of mathematics that was born from practical studies concerning lengths, areas and volumes in the early cultures, although it was not put into axiomatic form until the 3rd century BC. In this work, we will discuss the connection between these two timeless topics and show some ``new things in old things". There has been several works in this direction, but taking into account the didactic approach of the Enrico Fermi Summer School, we would like to address the subject and our audience in a new light.\\
{\bf PACS.} 02.40.Ma Global differential geometry. 42.50.-p Quantum optics. 32.81.-i Fiber optics.
\end{abstract}

\vspace{5mm}
\begin{center}
{\bf PROLOGUE}\\
\vspace{3mm}
{\it Two households, both alike in dignity,\\
In fair Varenna, where we lay our scene,\\
From ancient math break to new destiny,\\
Where bending light makes all inside unseen.\\
\vspace{3mm}
From forth the fertile loins of these two friends\\
A pair of ardent lovers take their lives:\\
Maxwell's fields and Einstein's geometric bends\\
Do with their union crown their parents' strifes.\\
\vspace{3mm}
The joyful passage of their deepest love,\\
And the continuance of their parents' fame,\\
Which, until the world shall end, nought can remove,\\
Is now the two hours' traffic of our game;\\
\vspace{3mm}
The which if you with patient ears attend,\\
What you have missed, our toil shall strive to mend.}\\
\end{center}

\vspace{5mm}
{\bf Adam.} As you know, optics is in some respect the foundation of geometry. Our perception of space is fundamentally based on sight, on the fact that light travels in straight lines almost always.

{\bf Ben.} But how can I see how light travels? 

{\bf Adam.} Fill the room with some smoke and then you see the light rays as straight, bright lines of light. Or sometimes, when the sun breaks through clouds, you see the rays as well --- or rather how their light is scattered in the mist.

{\bf Gabriel.} What do you mean with {\it almost always}? When is it not like that?

{\bf Adam.} Even if it sounds strange, light {\it can} travel in curved trajectories. Light can actually be bent. For light to be bent, the medium in which it travels must be inhomogeneous, its optical properties must vary in space.

{\bf Gabriel.} What exactly do you mean by optical properties? 

{\bf Adam.} We characterize the medium by its refractive index $n$ that is the ratio between the speed of light in vacuum and the speed of light in the medium. So in an inhomogeneous medium $n$ varies in space.

{\bf Ben.} But how does light travel when the refractive index varies? 

\begin{figure}
\centering
\includegraphics[scale=0.37]{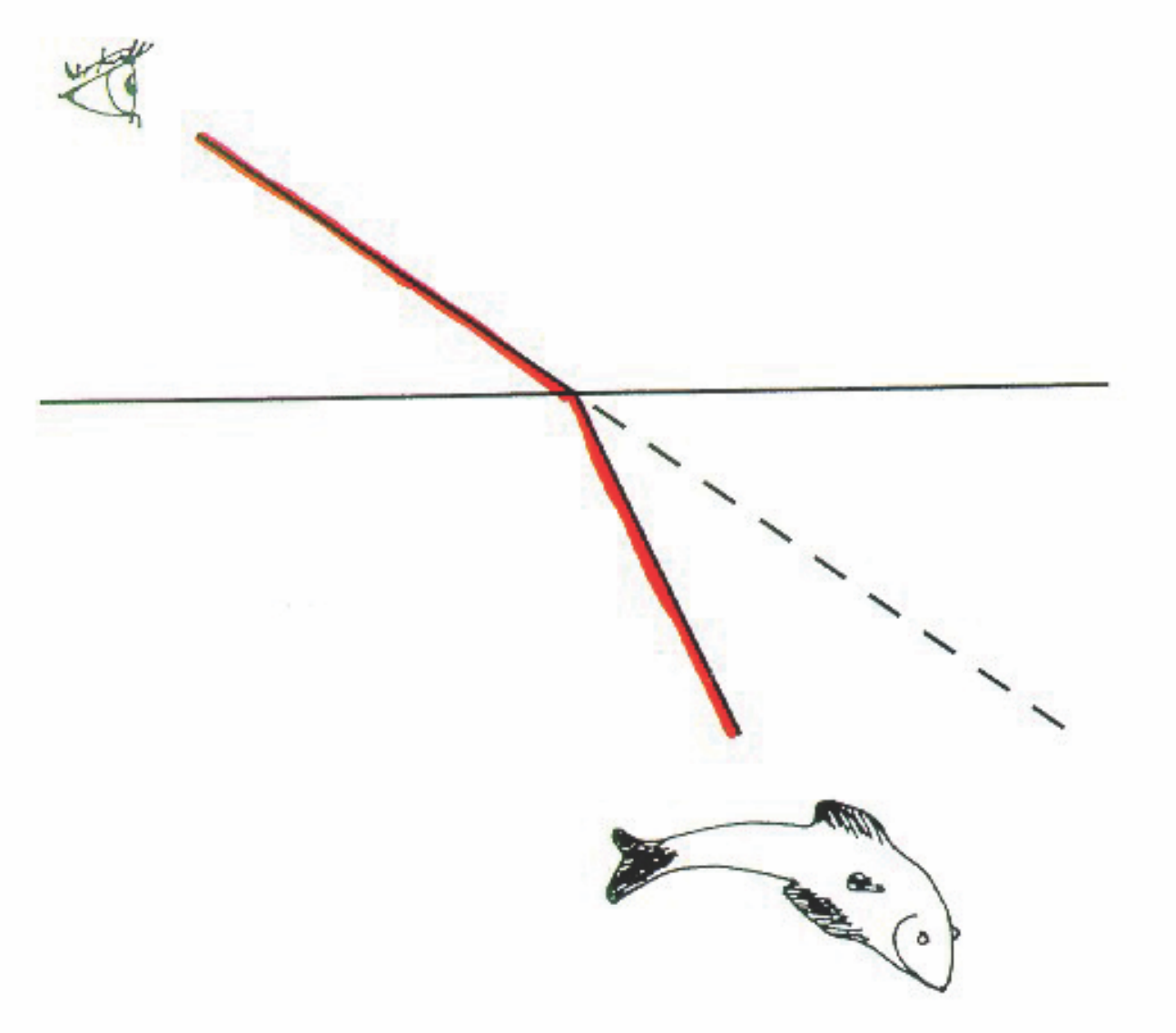}
\hspace{10mm}
\includegraphics[scale=0.55]{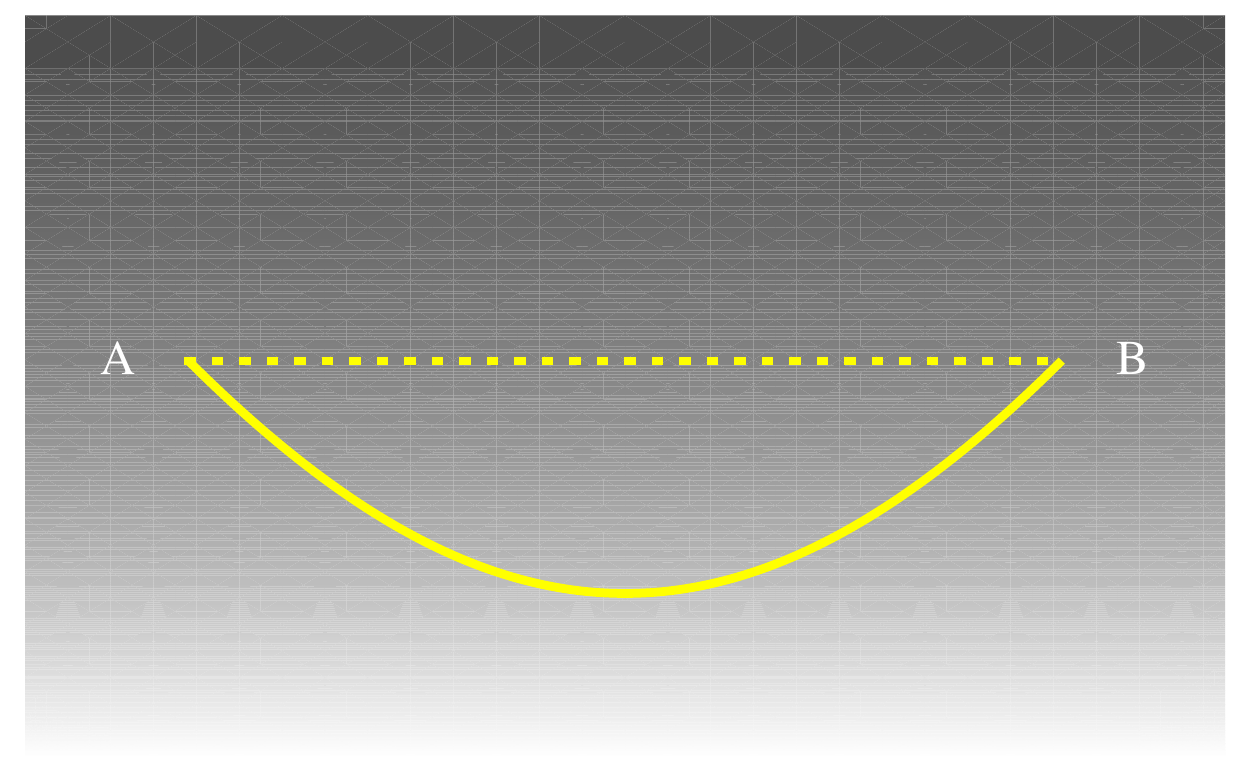}
\captionsetup{width=0.8\textwidth}
\caption{A fish inside a water tank has a {\it real} position and an apparent or {\it virtual} position (left). Also, the mirages produced in the desert are due to the gradient of temperature in the air which bend light (right).}
\label{fishmirage}
\end{figure}

{\bf Adam.} In 1662, Pierre de Fermat published what is now known as ``Fermat's principle", to explain how light travels from one place to another. For example, light coming from a fish inside a water tank gets refracted when it passes the interface between water and air (Figure~\ref{fishmirage}); when the medium varies, as at the interface between water and air, light is bent. A mirage in the desert is also caused by the bending of light: due to the temperature gradient in the air above the hot sand, the density of air varies a little bit, which leads to a small gradient of the refractive index that is enough, however, to bend light as a continuous series of refractions (Figure~\ref{fishmirage}). The result is astonishing: light appears to be reflected in hot air --- in the same way it is reflected at the surface of water. 

{\bf Ben.}  A traveller in the desert might be fooled to follow the mirage.

{\bf Adam.} Both the refraction of light at the water surface and the mirage are optical illusions: for our eyes objects --- like the fish --- appear at different positions from where they actually are.

{\bf Gabriel.} How did Fermat publish his principle in 1662? I do not think they had scientific journals at the time.

{\bf Adam.} In those years, scientific discoveries where published by sending letters to other colleagues explaining the discoveries. In this case, Fermat sent a letter to his friend, the physician of the King of France, who was also a science enthusiast. Then, this letter was sent to another colleague and so on, such that the scientific discoveries were spread slowly and only among an elite group of people. Also, Fermat used to tease his colleagues by sending them theorems without proof and challenging them to prove them. The most famous of Fermat's teasers is ``Fermat's Last Theorem", which took until the year 1995, when a proof was finally discovered. So that was quite a challenge.

{\bf Ben.} Was it the same with Fermat's principle?

{\bf Adam.} Well no, for a change, Fermat actually gave a proof; he used his principle to derive the law of refraction, Snell's Law, that was known at the time. He showed that Snell's Law and his newly discovered principle are consistent. As we can understand the bending of light in materials as a continuous succession of refractions, Fermat's Principle is proven (at least for a physicist). 

{\bf Gabriel.} You still have not explained the principle. What is it exactly?

{\bf Adam.} Fermat's Principle, also known as the principle of least time, states that the path taken by a light ray is an extremum of the optical path:
\begin{equation}
\text{Optical path length} = \int_A^B n\, \text{d} l \,.
\end{equation}
By the way, the concept of the refractive index $n$ was also invented by Fermat. He argued that an optical medium slows down light by a certain factor $n$, a bold statement, considering that this could not be measured at the time.

{\bf Gabriel.} By extremum you mean that it can be either a maximum or a minimum, right? A minimum would make sense --- it will minimize the time of travel like in other principles in classical mechanics. And, if you think about it, when light travels from $A$ to $B$, one could always find a longer path, so the maximum would not exist. The maximum does not make sense.

\begin{figure}
\centering
\includegraphics[scale=0.55]{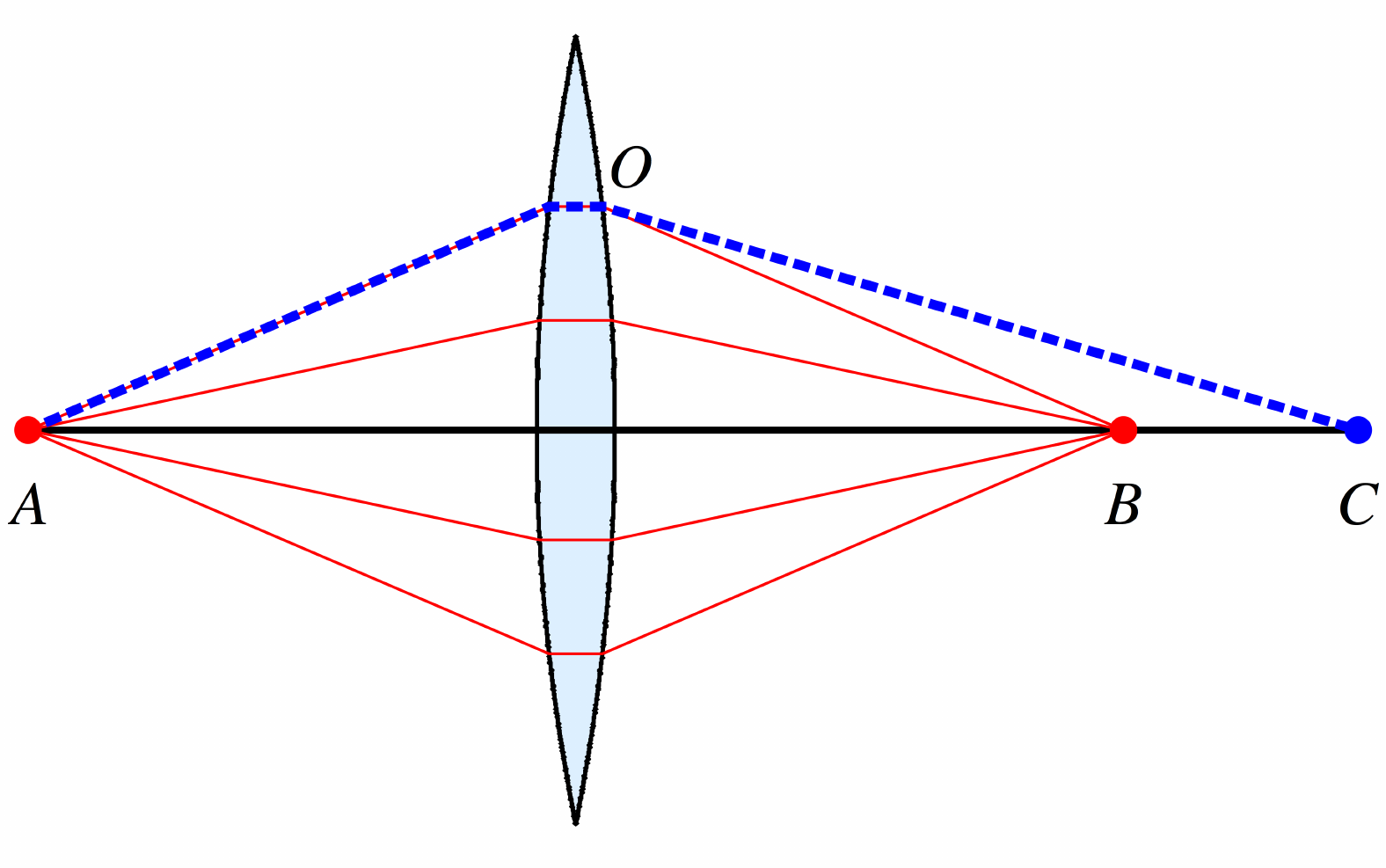}
\captionsetup{width=0.8\textwidth}
\caption{Focusing lens causes light rays to take a non minimal path from $A$ to $C$. The light rays take the path through $B$, but the optical path through $O$ is shorter.}
\label{lens}
\end{figure}

{\bf Adam.} Well, it does not always have to be a minimum. For example, take the case of a focusing lens, as illustrated in Figure~\ref{lens}. The natural path from $A$ to $C$ goes through $B$; this is how light travels. But there is an optically shorter path, which goes through $O$. You see this as follows. The lens focuses all light rays from $A$ in $B$. So, according to Fermat's Principle, the optical length of all these rays must be the same. (The lens gets thinner to compensate for the length difference.) Therefore, the optical length of the line $AB$ must be the same as the one of the polygon $AOB$. Now, the line $OC$ is shorter than the sum of $OB$ and $BC$. So the more complicated optical path $AOC$ is optically shorter than the direct path $AC$. 

{\bf Gabriel.} I understand that the optical path taken is not always a minimum, but it cannot be a maximum either --- it must be a saddle point in the multidimensional space of all paths! 

{\bf Ben.} But how can light figure out which path it should take? How does light know in advance where and how it should bend in order to arrive at a certain point and also at the shortest time? This does not make sense at all.

{\bf Adam.} Well, light rays do not know in advance where they end, but light is a wave. In principle, light goes into all directions and therefore can explore all possible paths.

{\bf Gabriel.} Logically, if the light from $A$ has not arrived at $B$, the principle does not apply, so there is no problem.

{\bf Ben.} I do not understand this.

{\bf Adam.} Fermat's Principle hints at something very deep. It has become one of the most influential principles in physics, if not the most influential one. It inspired Hamilton's principle in classical mechanics, the Principle of Least Action, and so Lagrangian and Hamiltonian dynamics. And it inspired quantum mechanics, because quantum mechanics can explain it.

{\bf Gabriel.} How does quantum mechanics explain Fermat's principle?

{\bf Adam.} According to quantum mechanics, light is associated with a wave function. The wave function explores all possible paths; each path contributes with a complex amplitude to the overall wave function. The sum of the amplitudes of all those paths eliminates most of them by destructive interference and gives rise to the path actually taken by constructive interference. This is the interpretation of Fermat's and Hamilton's principles by Richard Feynman --- another important insight in the long line of science inspired by Fermat's letter from 1662.

{\bf Ben.} But you want to discuss the connection of light and geometry.

{\bf Adam.} We have agreed that light does not always travel in straight lines and that this is explained by Fermat's principle. Let us make a little mathematical generalization. Let us write the optical path length as follows:
\begin{subequations}
\begin{align}
s = & \int_A^B n\,\text{d} l = \int_A^B n\sqrt{\text{d} x^2 +\text{d} y^2 + \text{d} z^2} =\int_A^B\sqrt{\sum_{i,j}g_{ij}\text{d} x^i\text{d} x^j} \,,\\
g_{ij}\,=&\, n^2 \delta_{ij}
\label{gij} 
\end{align}
\end{subequations}
where $\text{d} x^1=x$, $\text{d} x^2=y$, $\text{d} x^3=z $. The matrix $g_{ij}$ is known as the {\it metric tensor}. Therefore we have for the infinitesimal element $\text{d} s$ of the optical path length
\begin{equation}\label{pyta}
\text{d} s^2 = \sum_{i,j}g_{ij}\text{d} x^i\text{d} x^j \,.
\end{equation}
Here is where we begin to unite geometry and optics.

{\bf Ben.} How is this related to geometry?

{\bf Adam.} Well, the foundations of modern geometry lie exactly in this definition of the metric. Equation~(\ref{pyta}) describes a generalized Pythagorean relation; it gives the infinitesimal distance as a function of infinitesimal changes of the coordinates, like in the original Pythagoras Theorem $\text{d} s^2 = \text{d} x^2 + \text{d} y^2 + \text{d} z^2$. We get Pythagoras Theorem in the case of {\it flat space}, or {\it Euclidean space}, and only when Cartesian coordinates are used, as in this case we have $g_{ij}=\delta_{ij}$.

{\bf Gabriel.} So what happens if one has a different geometry? Curved space?

{\bf Adam.} Well, in this case, the functions $g_{ij}$ of Equation~(\ref{pyta}) will not be simply $\delta_{ij}$ as in the Euclidean case and no coordinate transformation could bring $g_{ij}$ to $\delta_{ij}$ for all points. 

{\bf Gabriel.} Understood. The metric tensor is simply a 3 $\times$ 3 matrix, isn't it?

{\bf Adam.} Yes, and the $x^i$ are the coordinates with the increments $\text{d} x^i$. For example, the Cartesian and spherical coordinates are
\begin{subequations}
\begin{alignat}{3}
x^1 = & \, x,		&\quad x^2=&\, y,			&\quad x^3=&\, z,\\
x^1 = & \, r, 		&\quad x^2=&\, \theta, 	&\quad	x^3=&\, \phi,
\end{alignat}
\end{subequations}
They are of course related to each other, as you know:
\begin{equation}
x = r\sin\theta\cos\phi\,,\quad y = r\sin\theta\sin\phi\,,\quad z = r\cos\theta\,.
\end{equation}

{\bf Gabriel.} Let me do a quick calculation. The Cartesian metric is $\text{d} s^2 = \text{d} x^2+\text{d} y^2+\text{d} z^2$ and from this I can calculate the metric in spherical coordinates by differentiation and algebraic simplification.

[After a short while.]

This is what I got:
\begin{subequations}
\begin{align}
\text{d} s^2 =&\, \text{d} x^2+\text{d} y^2+\text{d} z^2,\\
\text{d} s^2 =&\, \text{d} r^2+r^2 \text{d} \theta^2+r^2\sin^2 \theta \text{d} \phi^2.
\end{align}
\end{subequations}

{\bf Adam.} You see, both expressions for the metric have the same structure. You can read off the $g_{ij}$ for the spherical coordinates as
\begin{equation}
g_{ij}=\text{diag}\left(1,r^2,r^2\sin^2\theta\right).
\end{equation}
We can also define the metric tensor with raised indices $g^{ij}=(g_{ij})^{-1}$ as the inverse matrix of $g^{ij}$.

{\bf Ben.} Why do we need raised and lowered indices?

{\bf Adam.} Mathematically, they identify very different objects. The index notation is also very helpful clarifying expressions containing multiple summations over indices. This is called the Einstein's summation convention where we sum over repeated indices, one up and one down, so the summation sign can be omitted. For example 
\begin{equation}
g_{ij}\text{d} x^i\text{d} x^j \equiv \sum_{ij}g_{ij}\text{d} x^i\text{d} x^j \,.
\end{equation}
Einstein's summation convention is very convenient, but we shall write the sums explicitly, for not introducing too much new notation. 

{\bf Gabriel.} Is it always possible to find the inverse of the metric tensor?

{\bf Adam.} This is a good question and the answer is no. There are several examples of metrics that do not have a well defined inverse. One of them is the geometry of the invisibility cloak. I will discuss this soon.

{\bf Ben.} I am curious to hear about this one.

{\bf Adam.} As you see, there is an important difference between Cartesian and spherical coordinates: $g_{ij}$ is no longer proportional to the identity matrix, which is a general feature in coordinate transformations. We also need to define the determinant
\begin{equation}
g=\text{det}(g_{ij}).
\end{equation}

{\bf Ben.} Does the determinant of the metric tensor have a physical interpretation?

{\bf Adam.} It does: the infinitesimal volume element is given by
\begin{equation}
\text{d} V=\pm\sqrt{g}\,\text{d}^3 x
\label{volume}
\end{equation}
where the $\pm$-sign corresponds to right or left-handed coordinate systems. 

{\bf Ben.} Wait, why do we need the $\pm$ in definition (\ref{volume})?

{\bf Adam.} Let me show you an example and you will understand. Let us perform the following transformation on the original Cartesian coordinates: 
\begin{equation}
x'=x_0-x \,.
\label{foldtrans}
\end{equation}
In this case, $\text{d} x'=-\text{d} x$ and the metric remains the same, {\it i.e.},
\begin{equation}
\text{d} s^2=\text{d} x^2+\text{d} y^2+\text{d} z^2=\text{d} x'^2+\text{d} y^2+\text{d} z^2=\text{d} s'^2 \,.
\end{equation}
For the volume element we have
\begin{equation}
\text{d} V' = (-1) \text{d} x' \text{d} y' \text{d} z' = (-1)\,(-\text{d} x)\,\text{d} y\, \text{d} z=\text{d} x\, \text{d} y\, \text{d} z =\text{d} V\,.
\end{equation}

{\bf Ben.} I see, the  $\pm$-sign keeps the volume element positive, as it should be.

{\bf Gabriel.} What happens to the metric and its determinant after a coordinate transformation? 

{\bf Adam.} There is nothing particularly sophisticated. We have
\begin{subequations}
\begin{align}
\text{d} x^i =&\, \sum_{i'}\frac{\partial x^i}{\partial x'^{i'}}\,\text{d} x'^{i'}=\sum_{i'}\Lambda^i_{i'}\,\text{d} x'^{i'},\\
\text{d} x'^i =&\, \sum_{i}\frac{\partial x'^{i'}}{\partial x^i}\, \text{d} x^i=\sum_{i}\Lambda^{i'}_i \text{d} x^i\,.
\end{align}
\end{subequations}
This is how the differentials of the coordinates are transformed. From it also follows
\begin{equation}
\sum_{i'}\Lambda^i_{i'}\Lambda^{i'}_j=\delta^i_j.
\end{equation}
On the other hand, if we transform the derivatives we get:
\begin{subequations}
\begin{align}
\partial_i \equiv &\,\frac{\partial}{\partial x^i}=\sum_{i'}\frac{\partial x'^{i'}}{\partial x^i}\,\partial_{i'}'=\sum_{i'}\Lambda^{i'}_i\partial_{i'}'\,,\\
\partial_{i'}\equiv &\,\frac{\partial}{\partial x^{i'}}=\sum_{i}\frac{\partial x^{i}}{\partial x^{i'}}\,\partial_{i}=\sum_{i}\Lambda^{i}_{i'}\,\partial_{i}\,.
\end{align}
\end{subequations}
Notice that this is the exact opposite for the differentials; {\it i.e.}, differentials are transformed with $\Lambda^i_{i'}$ and the derivatives with $\Lambda^{i'}_i$. In the old literature, differentials and derivatives were known as contra-variant and co-variant vectors, although mathematicians do not approve of this anymore. They call them now vectors and one-forms, respectively. It is important that we discriminate between the two types of transformations. In general, anything with lower index behaves like a derivative and anything with upper index behaves like a differential. Moreover, they transform into each other through
\begin{equation}
F^i=\sum_j g^{ij}F_j\,, \quad F_i=\sum_j g_{ij} F^j \,.
\end{equation}

{\bf Gabriel.} So any $F_i$ transforms like $\partial_i$ and any $F^i$ transforms like $\text{d} x^i$. Can you give us a physical example of these quantities?

{\bf Adam.} Well, for example the electrostatic field can be derived from the electric potential $U$ as $E_{i}=\partial_{i}U$, therefore is a derivative, and so it must transform like a derivative.  Another example is the momentum. There are two types of momenta: The {\it canonical momentum}, which is the gradient of the action and therefore a one-form, and the {\it dynamical momentum} which is proportional to the velocity, and therefore a vector. The connection between them is a geometrical transformation between a one-form and a vector. The two momenta generally do not agree.

{\bf Ben.} Seriously? When do they disagree?

{\bf Adam.} For the case of light in a medium. The {\it Minkowski momentum} is given by the derivative of the action $S$ as
\begin{equation}
p_i=\partial_i S=\hbar\,\partial_i \varphi\,,
\end{equation}
where $\varphi$ is the phase of the light field. As $\varphi$  is the integral of the wavenumber $k$ over the trajectory, and $k=n\,\omega/c$, we obtain
\begin{equation}
p=n\, \frac{\hbar\omega}{c} \,.
\end{equation}
This, surely, is a natural way of defining the momentum for the light wave associated with a single photon. On the other hand, there is a competing expression for the momentum called the {\it Abraham momentum} that comes from the particle picture. From Einstein's $E=mc^2$ and Planck's $E=\hbar\omega$ follows that the dynamical mass of a photon is $m=\hbar\omega/c^2$. For the momentum $p=mv$ with $v=c/n$ we thus obtain
\begin{equation}
p=\frac{\hbar k}{n}=\frac{\hbar \omega}{nc} \,.
\end{equation}

{\bf Gabriel.} Oh I see the contradiction now. The two momenta differ by $n^2$ and they can only coincide when $n=1$, {\it i.e.}, in vacuum. In all other cases they are different.

{\bf Adam.} There has been a debate for over a century about the question which momentum is the correct one. 

{\bf Ben.} Let us not waste our time with endless discussions. You promised to tell us something important about geometry and light.

{\bf Adam.} So far we have talked about ray optics and some concepts of geometry. Now, let us move on to proper electromagnetism in media and describe its connection with geometry. Let us begin at the beginning, at Maxwell's equations in free space:
\begin{equation}
\nabla \cdot \bm{E}=0, \quad \nabla \times \bm{E}= -\partial_t \bm{B}, \quad
\nabla \cdot \bm{B}=0, \quad \nabla \times \bm{B}= \frac{1}{c^2}\partial_t \bm{E} \,.
\label{maxwell}
\end{equation}
Imagine that space is curved or expressed in curved coordinates. To describe the effect of the geometry on electromagnetic fields, we need to express the mathematical operations that appear in Maxwell equations in terms of general coordinates and the metric $g_{ij}$. We need to formulate the divergence and the curl in general coordinates and curved geometries. For this we borrow a few results from differential geometry. For the divergence we have
\begin{equation}
\nabla \cdot \bm{E} =\frac{1}{\sqrt{g}}\sum_{ij}\partial_i \sqrt{g} g^{ij}E_j\,.
\label{divE}
\end{equation}

{\bf Ben.} Why do you include the $\sqrt{g}$ inside and outside of the sum, do they not cancel out?

{\bf Gabriel.} They do not cancel each other, because the second $\sqrt{g}$ is inside a derivative $\partial_i$. So in general, if $g$ depends on space they cannot cancel out.

{\bf Adam.} For a Cartesian system we get the usual $\nabla\cdot\bm{E}=\partial_xE_x+\partial_yE_y+\partial_zE_z$. For any other coordinate system it is more complicated.

{\bf Gabriel.} Let me do the calculation for the spherical coordinate system:
\begin{equation}
g_{ij}=\text{diag} (1,r^2,r^2\sin^2\theta)\,, \quad g^{ij}=\text{diag} \left(1,\frac{1}{r^{2}},\frac{1}{r^{2}\sin^{2}\theta}\right),\quad g=r^4\sin^2\theta \,.
\end{equation}
So, if we take the electric field in spherical coordinates and calculate its divergence, we get
\begin{equation}
\nabla \cdot \bm{E} = \frac{1}{r^2 \sin \theta}\left(\partial_r r^2\sin\theta\, E_r+\partial_\theta\sin\theta E_\theta+\partial_\phi \frac{1}{\sin\theta} E_\phi\right).
\end{equation}
If we consider a radially-symmetric field where $E_\theta=E_\phi=0$ we have
\begin{equation}
\nabla \cdot \bm{E} = \partial_r E_r+\frac{2}{r}E_r 
\end{equation}

{\bf Adam.} --- which is the divergence in spherical coordinates you probably know.

{\bf Gabriel.} I remember its tedious derivation. So my calculation shows that formula (\ref{divE}) is a very effective way for obtaining the divergence. But what about the curl?

{\bf Adam.} For the curl we get from differential geometry
\begin{equation}
(\nabla \times \bm{E})^i = \pm \sum_{jk}\frac{[ijk]}{\sqrt{g}}\partial_j E_k
\label{curlE}
\end{equation}
where $[ijk]$ is the total antisymmetric symbol defined by $[123]=1$ and $[jik]=-[ijk]$. The $\pm$ sign is to account for right/left-handed coordinate systems, just as we did for the volume element. Formula (\ref{curlE}) coincides with the usual definition of the curl in right-handed Cartesian coordinates as
\begin{equation}
(\nabla \times \bm{E})^x = (\nabla \times \bm{E})^1 = \partial_2 E_3-\partial_3 E_2 = \partial_y E_z-\partial_z E_y.
\end{equation}

{\bf Ben.} Is the total antisymmetric symbol a $3\times 3 \times 3$ tensor?

{\bf Adam.} Not really, it is more like a cube of numbers that depend on the indices; it does not fulfil the definition of a tensor, it does not transform like a tensor.

{\bf Gabriel.} In formula (\ref{curlE}) the metric appears only in the determinant. What does this mean?

{\bf Adam.} It means that the value of the curl depends only on some information encoded in the metric; it depends only on the infinitesimal volume and not on the shape of the volume element. Now, I would like to show you how to express Maxwell's equations in generalized coordinates:
\begin{equation}
0=\nabla \cdot \bm{E} = \frac{1}{\sqrt{g}}\sum_{ij}\partial_i \sqrt{g} g^{ij}E_j
\end{equation}
that we can write as
\begin{equation}
\sum_i \partial_i D^i = 0 
\end{equation}
if we define a new field $D^i$ as
\begin{equation}
\label{defD}
D^i\equiv\pm \varepsilon_0 \sum_{j} \sqrt{g} g^{ij}E_j \,.
\end{equation}
Let us regard the field $D^i$ as the dielectric displacement $\bm{D}$.

{\bf Gabriel.} What about the usual connection between $\bm{D}$ and $\bm{E}$?

{\bf Adam.} It still holds; if we define the electric permittivity as
\begin{equation}
\varepsilon^{ij}=\pm \sqrt{g}g^{ij}
\label{permit}
\end{equation}
we get the constitutive equation
\begin{equation}
\bm{D}=\varepsilon_0 \varepsilon \bm{E} \,.
\end{equation}

{\bf Ben.} The dielectric displacement also enters another Maxwell equation, $\nabla\times\bm{H}=\partial_t \bm{D}$. Is your definition consistent there?

{\bf Adam.}  Let us check. We get from Maxwell's equations (\ref{maxwell}) and expression (\ref{curlE})
\begin{subequations}
\begin{alignat}{2}
\pm \sum_{jk}\frac{[ijk]}{\sqrt{g}}\partial_j B_k =&\,\frac{1}{c^2}\partial_t \sum_j g^{ij}E_j\,, \quad & \text{and so\hspace{7.5mm}} \\
\pm \sum_{jk}[ijk]\partial_j B_k =&\,\frac{1}{c^2}\partial_t \sum_j \left(\sqrt{g}g^{ij}\right)E_j\,, \quad & \text{which gives} \\
(\nabla \times \bm{B})^i =&\,\frac{1}{c^2\varepsilon_0}\partial_t D^i. \quad & \quad
\end{alignat}
\end{subequations}

{\bf Gabriel.} For expressing this as a Maxwell equation, we need to define the $\bm{H}$ field.

{\bf Adam.} Yes, if we use $H_i\equiv\varepsilon_0 c^2 B_i$ then all the equations match
\begin{equation}
\sum_{jk}[ijk] \partial_j H_k = \partial_t D^i.
\end{equation}

{\bf Ben.} --- which is $\nabla\times\bm{H}=\partial_t \bm{D}$.

{\bf Adam.} Now you probably believe that we can express the other two of Maxwell's equations in a neat form if we assume a similar relationship between $\bm{H}$ and $\bm{B}$:
\begin{equation}\label{defB}
B^i=\pm \mu_0 \sum_{j} \sqrt{g} g^{ij}H_j \,,
\end{equation}
or, expressed as a constitutive equation,
\begin{equation}
\bm{B}=\mu_0 \mu \bm{H}
\end{equation}
with the magnetic permeability
\begin{equation}
\mu^{ij}=\pm \sqrt{g}g^{ij} \,.
\label{permea}
\end{equation}
We obtain the macroscopic Maxwell equations in Cartesian coordinates
\begin{equation}
\nabla \cdot \bm{D}=0, \quad \nabla \times \bm{E}= -\partial_t \bm{B}, \quad
\nabla \cdot \bm{B}=0, \quad \nabla \times \bm{H}= \partial_t \bm{D} \,.
\label{macromax}
\end{equation}
The geometry appears like a medium with electric permittivity (\ref{permit}) and magnetic permeability (\ref{permea}).

{\bf Gabriel.} But wait, if $H_i=\varepsilon_0 c^2 B_i$ then $B^i$ and $B_i$ satisfy $B^i=\sum_{j} \sqrt{g}g^{ij}B_j$, which is not how we raise and lower indices. 

{\bf Adam.} I misused the index notation a bit. It is important to understand that $B_i$ and $B^i$ do not live in the same space, $B_i$ lives in the original general geometry where one raises and lowers indices with $g^{ij}$, while $B^i$ lives in a flat, medium-filled space in Cartesian coordinates. We could introduce curved coordinates there as well and then the lowering and raising of indices is done with the corresponding metric tensor.

{\bf Gabriel.} This juggling of coordinates seems very confusing. 

{\bf Adam.} I can offer an alternative derivation of the connection between geometries and media. You know that we can derive field equations, including Maxwell's equations, from an action principle. The fields minimize the action $S$ that is given by an integral of a specific functional, the Lagrangian density, over space and time. For electromagnetic fields, we need to express the fields via the electromagnetic potentials. Let us put the fields $\bm{E}$ and $\bm{B}$ in terms of the magnetic vector potential $\bm{A}$ in the Coulomb gauge, {\it i.e.},
\begin{equation}
\bm{E}=-\partial_t \bm{A}\,,\quad \bm{B}=\nabla \times \bm{A}\,,
\end{equation}
or in index notation
\begin{equation}
E_i=-\partial_t A_i\,,\quad B^i= \pm\sum_{jk}\frac{[ijk]}{\sqrt{g}}\partial_j A_k \,.
\end{equation}
It is known \cite{jackson} that the action of the electromagnetic field is 
\begin{equation}\label{action}
S=\frac{\varepsilon_0}{2}\int (\bm{E}^2-c^2\bm{B}^2)\,\text{d} V \text{d} t \,.
\end{equation}

{\bf Gabriel.} Let me check.

[after a while]

Yes I see that from the Euler-Lagrange equations \cite{jackson} follow Maxwell's equations (\ref{maxwell}).

{\bf Adam.} Now, let us define a new set of fields as
\begin{equation}
E_i\equiv -\partial_t A_i\,,\quad B^i\equiv \sum_{jk} [ijk]\partial_j A_k\,,
\end{equation}
{\it i.e.}, the same $E_i$ but a new $B^i$, although we keep the same symbol $B^i$. Let me re-express the action (\ref{action}) in terms of the new fields:
\begin{equation}
S=\pm \frac{\varepsilon_0}{2}\int \sum_{ij} \left(g^{ij}E_i E_j-\frac{c^2}{g}g_{ij}B^iB^j\right)\sqrt{g}\,\text{d}^3 x\, \text{d} t \,.
\end{equation}
I use that $g^{-1/2}g_{ij}$ is the inverse matrix of $g^{1/2}g^{ij}$ and follow the definitions of $D^i$ and $B^i$ given in Eqs.~(\ref{defD}) and (\ref{defB}). In this way I obtain the following action
\begin{equation}
S=\frac{1}{2}\int  (\bm{E}\cdot \bm{D}-\bm{B}\cdot\bm{H})\,\text{d} ^3 x \,\text{d} t\,,
\end{equation}
which is the usual action that leads to the macroscopic Maxwell's equations (\ref{macromax}). Do you see how short and elegant this derivation is?

{\bf Ben.} Now we have two derivations, but both are hard to swallow.

{\bf Adam.} We started from Maxwell's equations in empty space --- without any media, but equipped with a metric, a geometry that might be given by a coordinate transformation of flat space or by genuinely curved space. We defined new fields where we expressed the geometry in terms of the constitutive equations of a medium. The new fields obey the macroscopic Maxwell equations in Cartesian coordinates. So we have a choice --- we can either understand the geometry as a geometry or as a medium. For the medium, we got the relations
\begin{equation}
\label{epmu}
\varepsilon^{ij}=\mu^{ij}=\pm\sqrt{g} g^{ij},
\end{equation}
which is one of the main results of transformation optics. 

{\bf Ben.} I see that permittivity and permeability tensors are related to the metric tensor.

{\bf Gabriel.} The values of $\varepsilon$ and $\mu$ describe a medium, more precisely, the electromagnetic properties of a material, and the metric tensor describe the coordinate system. 

{\bf Adam.} Not only a coordinate system. In general, the metric defines a spatial geometry that is curved, {\it i.e.} it cannot be reduced to Cartesian flat space by coordinate transformations. Equation~(\ref{epmu}) describes the connection between medium and geometry.

{\bf Ben.} We also got that  $\varepsilon^{ij}=\mu^{ij}$. What does this mean physically?

{\bf Adam.} A medium that satisfies this relationship is called an {\it impedance-matched} medium (impedance-matched with the vacuum).

{\bf Ben.} I know from electrical engineering that impedance-matched transmission lines do not reflect. So these media do not reflect light?

{\bf Adam.} Almost. For one-dimensional systems, like transmission lines or cables, it is straightforward to show that there are no reflections in such media. Let me show you. We can obtain the wave equation for a one-dimensional medium from the action (\ref{action}) where in 1D we have
\begin{equation}
E=-\partial_t A\,,\quad B=\partial_z A\,,
\end{equation}
Assume the constitutive equations $D=\varepsilon_0\varepsilon E$ and $B=\mu H/(\varepsilon_0c^2)$ with scalar $\varepsilon$ and $\mu$. We get the action
\begin{equation}
S=\frac{\varepsilon_0}{2}\int \left(\varepsilon(\partial_t A)^2-\frac{c^2}{\mu}(\partial_z A)^2\right)\text{d} x\,\text{d} t \,,
\label{action0}
\end{equation}
which leads to the wave equation
\begin{equation}
\frac{1}{\varepsilon} \partial_z \frac{1}{\mu}\partial_z A-\frac{1}{c^2}\partial_t^2 A=0 \,.
\end{equation}
In an impedance-matched material $\varepsilon=\mu=n$, so
\begin{equation}
\frac{1}{n}\partial_z\frac{1}{n}\partial_z A - \frac{1}{c^2}\partial_t^2 A=0 \,.
\end{equation}
Now we perform a simple coordinate transformation,
\begin{equation}
s=\int n\, \text{d} z
\end{equation}
and get the following equation
\begin{equation}
\left(\partial_s^2-\frac{1}{c^2}\partial_t^2\right) A=0 \,.
\end{equation}
What we have obtained here is the wave equation in free space where there is no scattering. This proves that impedance-matched media in 1D --- cables, transmission lines, the lot --- do not reflect.

{\bf Gabriel.} What happens in more than one dimension?

{\bf Adam.} In higher dimensions, this is no longer strictly true. The reason is the following. For two or more dimensions the corresponding geometry can contain curvature, which may cause scattering and hence reflections. One-dimensional geometries, on the other hand, are always flat and therefore cannot reflect.

{\bf Gabriel.} What we have done here is to start from Maxwell's equation in a certain geometry and, as we realised, we connected the geometry with a certain medium. Can it be done the other way round?

{\bf Adam.} Yes, an impedance-matched medium always corresponds to a geometry, because for a given $\varepsilon$ we can calculate the corresponding metric tensor. For this we need to compute the determinant
\begin{equation}
\text{det}(\varepsilon^{ij})=\pm g^{3/2}\frac{1}{g}=\pm \sqrt{g} \,.
\end{equation}
Now we obtain from relations (\ref{epmu})
\begin{equation}
g^{ij}=\frac{\varepsilon^{ij}}{\text{det}(\varepsilon^{ij})}\,.
\end{equation}
A spatial geometry appears to light as an impedance-matched medium, and an impedance-matched medium appears as a geometry: an impedance-matched medium {\it is} a geometry.

{\bf Ben.} Interesting, but what can we do with this theory? It seems very difficult to make impedance-matched materials in 3D.

{\bf Adam.} There are some cases where you can get around impedance-matching. Suppose that your field has a fixed polarization. Then you only need to impose the relations between geometry and media for the relevant components of the $\varepsilon$ and $\mu$ tensors, which tremendously simplifies things. Or, suppose that you need $\varepsilon=\mu$ where $\varepsilon$ is proportional to the unity matrix. In this case the material is isotropic. As long as you are concerned only about wave propagation within the regime of geometrical optics, you can characterize the medium solely by the refractive index profile, {\it i.e.} by the square root of the product of $\varepsilon$ and $\mu$. Therefore it does not matter if one modulates $\varepsilon$ or $\mu$, what counts is the product of the two.

{\bf Ben.} What do you mean by geometrical optics? Ray optics?

{\bf Adam.} No, geometrical optics is an approximative regime of electromagnetic waves. It says that the waves are guided by their phases, and the phase fronts $\varphi$ are governed by the dispersion relation
\begin{equation}
(\nabla\varphi)^2 = k^2 = n^2\frac{\omega^2}{c^2} \,,
\end{equation}
also known as the eikonal equation. This regime is valid when 
\begin{equation}
|\nabla \lambda| \ll 1
\end{equation}
for the spatially-dependent wavelength 
\begin{equation}
\quad \lambda = \frac{2\pi c}{n(\bm{x})\omega} \,,
\end{equation}
{\it i.e.} when the wavelength does not vary much. Geometrical optics is violated at sharp interfaces, for example. Light rays are the lines orthogonal to the phase fronts of the light waves.

{\bf Gabriel.} But what is really new here? We have been talking about mathematics, classical electromagnetism, and geometry. Scientists have been doing this for at least a hundred and fifty years, since the time of the discovery of Maxwell's equations. 

{\bf Ben.} Do we not just make simple things difficult?

{\bf Adam.} Well, there is nothing new under the sun, but this does not mean that everything is already discovered, quite the opposite. With fresh eyes we can look at something old and find something new, as Sir Michael Berry said: ``{\it ... to find new things in old things.}" 

{\bf Ben.} So how can we take these ideas from geometry and find new things?

{\bf Adam.} By encoding the functionality of a device in a geometrical concept and translating the geometry into a medium that does it in practice. 

{\bf Ben.} Can you give an example?

{\bf Gabriel.} What kind of geometries encode useful functionalities?

{\bf Adam.} It turns out that some of the most interesting geometries for optical applications are those that represent coordinate transformations.

{\bf Gabriel.} But how can anything interesting can come out of a coordinate transformation? Are not physical phenomena supposed to be independent of the choice of coordinate systems? If a medium implements just a coordinate transformation nothing would change for light traveling through.

\begin{figure}[h]
\centering
\includegraphics[scale=0.8]{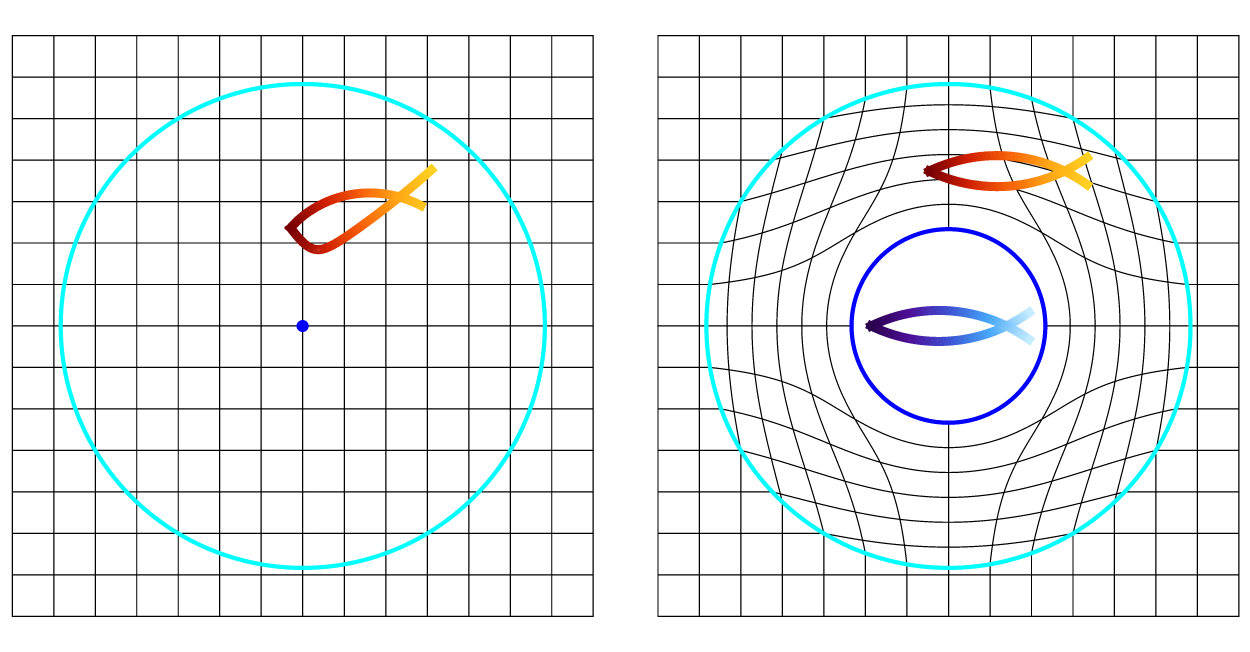}
\captionsetup{width=0.8\textwidth}
\caption{Virtual and physical space for the cloaking device.}
\label{cloakingfigure}
\end{figure}

{\bf Adam.} This is exactly the point of invisibility. There you want to disguise the fact that light is traveling through a medium. Let me explain you this in detail. Consider a coordinate transformation of space (that we can implement with a medium according to the recipe that we just described). In the case of a coordinate transformation, it is instructive to think of two spaces: the {\it physical space}, which is physical reality where light may be bent by a medium and the so-called {\it virtual, electromagnetic, or optical space} which describes what we see and where light travels in straight lines. Have a look at Figure \ref{cloakingfigure}. 

In virtual space, I draw a straight Cartesian grid of orthogonal coordinate lines, in physical space they are deformed by a coordinate transformation. The transformation shown in the picture is the following. Pick a point in virtual space and expand it to a finite size in physical space. You may do this in a certain region in space that represents the actual device. The shape of this region is not particularly important, so let us assume it is a sphere surrounding the chosen point. 

{\bf Ben.} But how does this construction make things invisible?

{\bf Adam.} Physical space represents physical reality and virtual space the optical illusion the device creates. There an extended region in physical space has been demagnified to the size of a single point. As a point is invisible the entire region of physical space surrounded by the device, and everything inside it, has become invisible, too. Moreover, as the device is just made by a coordinate transformation of empty space one would never notice that it is there; it is as invisible as empty space, creating the ultimate optical illusion, perfect invisibility.

{\bf Ben.} Let us not get carried away. How does this cloaking effect happen in practice, what is the actual path taken by the light?

{\bf Adam.} I have drawn the path of a couple of light rays in the Figure \ref{cloakingfigure}. In virtual space they are straight lines, as the geometry is flat and Cartesian there. In physical space these lines are simply transformed by the coordinate transformation. You see how they bend around the hidden interior of the device. So the transformation describes how invisibility is achieved by light bending.

{\bf Gabriel.} What about the central ray you drew, the ray that goes directly into the invisible point in virtual space? It appears it follows an impossible trajectory, as it suddenly changes direction, and also the trajectory is not unique.

{\bf Ben.} So there is a problem with this cloaking device --- as I suspected.

{\bf Adam.} Realistically, light would not consist of just a single ray, but a beam in which one ray is of no great importance. But there still is a problem with cloaking. Remember that what happens in physical reality happens in physical space at the same time. The two spaces are just connected by a spatial transformation. Imagine you follow the central light ray in virtual space. It passes through ---- goes around --- the invisible point. In physical space it must go around the invisible region, and it must do so during the same time it takes for light to pass a single point. 

{\bf Ben.} In no time at all, which is impossible. The speed of light must go to infinity at the cloaking device.

{\bf Adam.} Yes, this is why perfect cloaking is not possible, but imperfect cloaking is.

{\bf Ben.} Suppose that perfect cloaking were possible and I could sit inside a cloaking device, would I see something?

{\bf Gabriel.} No, if you cannot be seen you will not see either.

{\bf Adam.} Not so fast. Remember that what cloaking does is demagnifying you to the size of point. So the real question is: could a point see something? 

{\bf Ben.} Not very much, but it could see whether it is bright or dark.

{\bf Adam.} Yes, cloaking could be used for sensing, for making sensors that perturb the electromagnetic field as little as possible \cite{Uhlmann}.

{\bf Gabriel.} Have cloaking devices ever been made?

{\bf Adam.} They have been achieved for microwaves \cite{Schurig}. It is very hard to make them for visible light. You can see why in Figure \ref{cloakingfigure}. The orthogonal grid in virtual space has turned into a skewed grid in physical space, which means that the material implementing the transformation must be highly anisotropic. Such a material can be made for microwaves --- within limits, and it has been made \cite{Schurig}, but only for one polarization where not all components of the dielectric tensors matter, and within the limit of geometrical optics where impedance matching can be relaxed. 

{\bf Ben.} Impressive, but what about the impossibility of perfect cloaking? Here we have a device and it shows that cloaking works.

{\bf Adam.} It only works for a single frequency. In the monochromatic limit you can reach an infinite phase velocity of light, but you cannot transmit information, as information implies finite bandwidth. Cloaking is much easier for sound waves \cite{Acoustics,Acoustics2,Acoustics3}, for several reasons, practical and more fundamental ones. A practical reason is that the wavelength of sound is on the human scale, so acoustical materials can easily be structured. Furthermore, the speed of sound can vary a great deal in different materials, so extreme acoustical media are easier to make with composite materials than extreme optical media. There is also a fundamental reason why acoustical cloaking is easier: sound waves are not as constrained by relativistic causality as light --- the speed of sound is significantly smaller than the speed of light in vacuum. So it is possible to have a very large speed of sound over a wide frequency window. It is even possible to achieve cloaking of earth quakes.

{\bf Ben.} Now you are definitely spinning a yarn.

{\bf Adam.} Not quite, earth quakes are deformation waves and the most dangerous ones propagate on the surface. They could, in principle, be guided around objects one wishes to protect in the same way light is guided around the hidden object in the cloaking device. There are serious people working on these applications that have already performed some experiments being performed \cite{EarthquakePRL}.

{\bf Ben.} But let us go back to optics where cloaking is difficult. Is there any way to make it easier?

\begin{figure}
\centering
\includegraphics[scale=0.7]{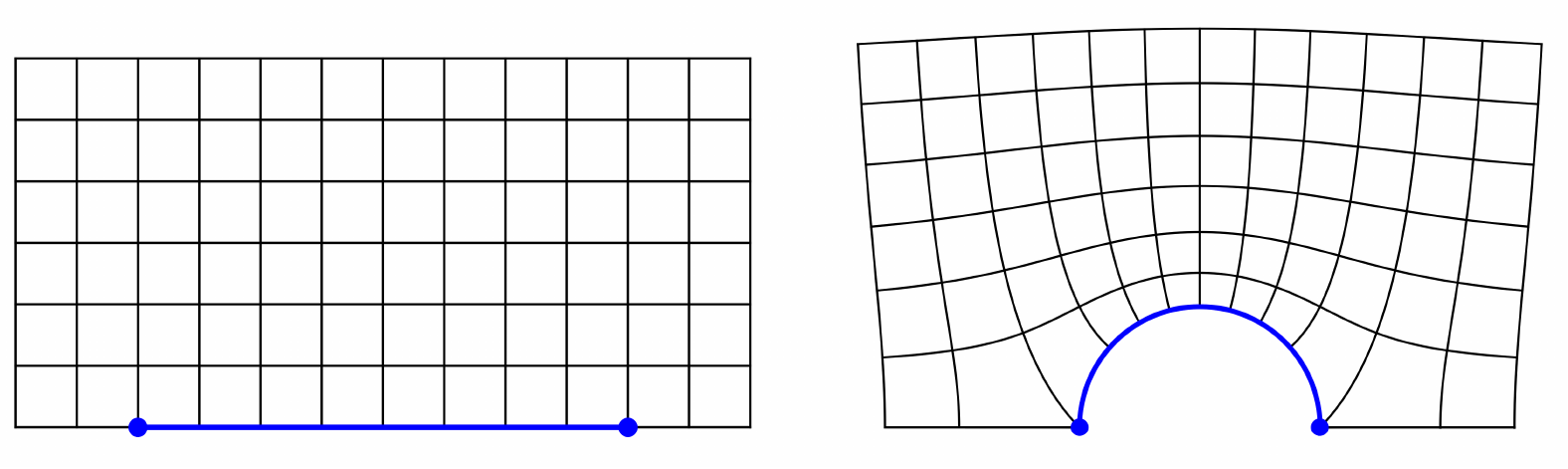}
\captionsetup{width=0.8\textwidth}
\caption{Carpet cloaking hiding an object inside the circle shown in the right hand side (physical space). Light traveling between the two gray points will surround the circle.}
\label{carpet}
\end{figure}

{\bf Adam.} You can make cloaking easier by being less ambitious, for example in what is called {\it carpet cloaking} \cite{Carpet}. Have a look at Figure \ref{carpet}. Here you see virtual space and physical space as before, but the transformation connecting the two spaces preserves the orthogonality of the grid of coordinates: it is conformal. With a conformal transformation you cannot turn a finite region into a single point, but you can make a curved line flat. With a quasi-conformal transformation \cite{Carpet} you can make a 3D objects appear flat \cite{Wegener}. You can put something under the surface --- under the carpet --- without creating a bump. 

{\bf Ben.} What else can we do with transformation optics, apart from cloaking?

\begin{figure}
\centering
\includegraphics[scale=0.5]{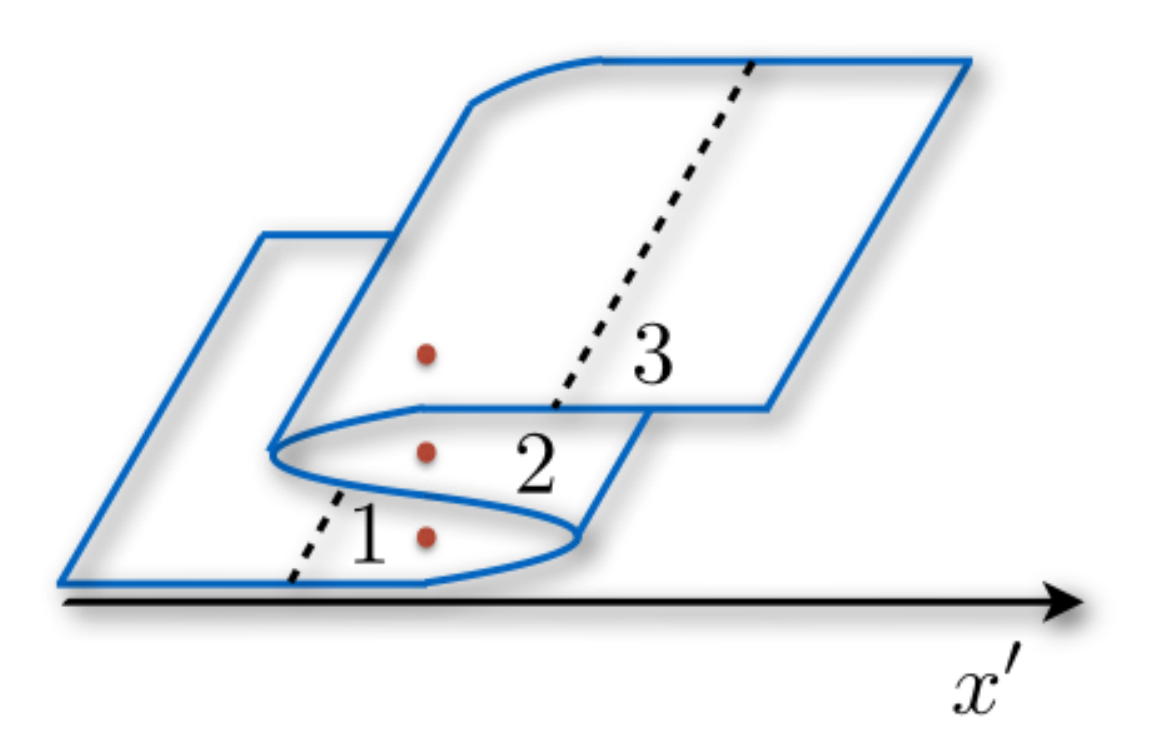}\hspace{10mm}
\includegraphics[scale=0.5]{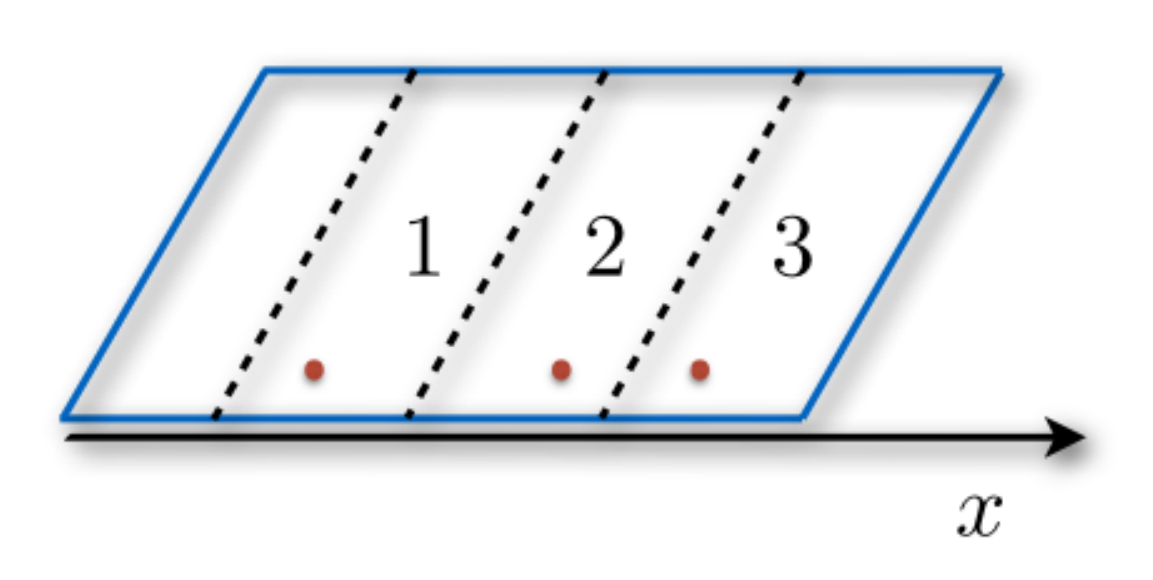}
\captionsetup{width=0.8\textwidth}
\caption{Folding in virtual space (left) and diagram for the transformation from virtual space $x'$ to real space $x$ (right).}
\label{folding}
\end{figure}

{\bf Adam.} Another example is {\it perfect imaging} based on the concept of {\it negative refraction}. Imagine that you represent physical space by a stack of planes, like a stack of sheets of paper. Take each sheet and fold it as shown in Figure \ref{folding}. Suppose that the folded sheet represents virtual space, whereas the original sheet is a sheet of physical space. They are not the same: on the fold three patches of physical space are made indistinguishable. There, any point in virtual space corresponds to three points in physical space. The important point is that the the electromagnetic field just perceives virtual space, but our three patches of physical space are identical in virtual space. From this follows that electromagnetic field must be absolutely identical in the three different regions of physical space. Therefore, any device that performs such a transformation would act like a perfect copy machine of the electromagnetic field, {\it i.e.}, it would make a {\it perfect lens} \cite{Pendry}.

{\bf Gabriel.} How do you describe the folding transformation mathematically?

\begin{figure}
\centering
\includegraphics[scale=0.5]{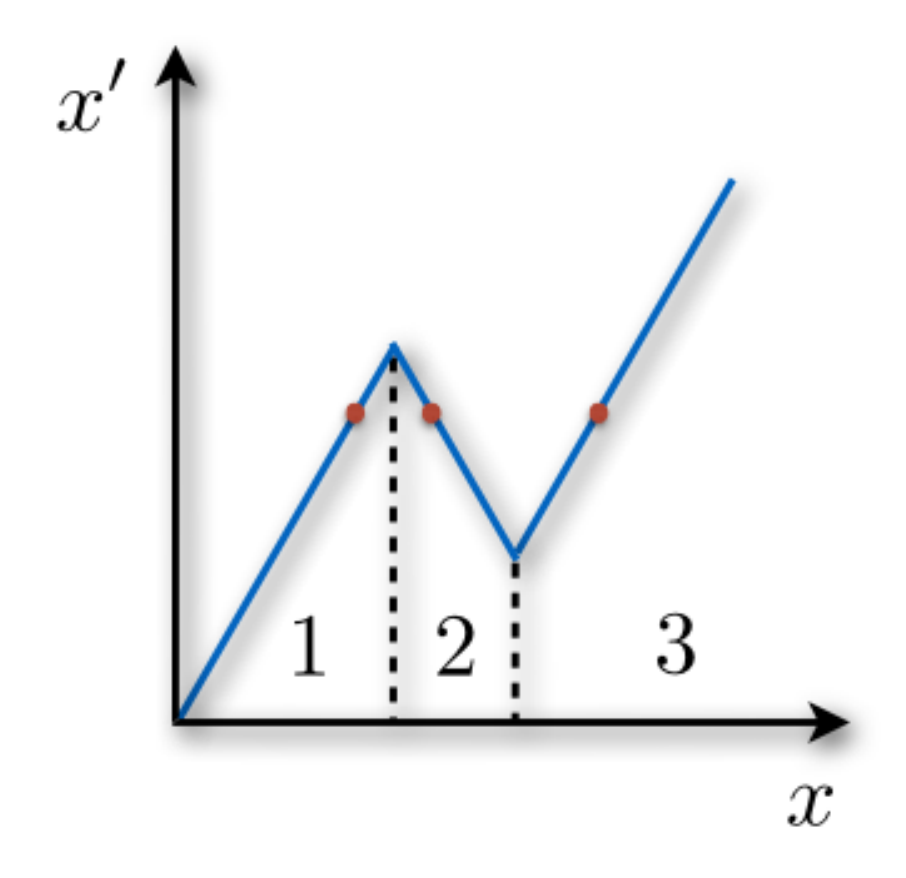}
\captionsetup{width=0.8\textwidth}
\caption{Folding transformation from virtual to real space $x'\rightarrow x$.}
\label{diagram}
\end{figure}

{\bf Adam.} Let me describe this with a diagram (Figure \ref{diagram}). Here $x'$ is a coordinate in virtual space and $x$ the transformed coordinate in physical space. They are connected by formula (\ref{foldtrans}). We see that the slope of the curve in region 2 is $\text{d} x'/\text{d} x=-1$. If we look at the metric
\begin{equation}
\text{d} s^2=\text{d} x'^2+\text{d} y'^2+\text{d} z'^2=\left(\frac{\text{d} x'}{\text{d} x}\right)^2\text{d} x^2+\text{d} y^2+\text{d} z^2\,,
\end{equation}
we see that the prefactor in front of $\text{d} x^2$ is the square of $-1$, hence the metric is the same.

{\bf Ben.} So, if there is no change in the geometry, why would there be any optical effect at all? Something must be wrong.

{\bf Gabriel.} I remember the discussion about the sign of the volume element (\ref{volume}). There we argued that if we reverse the direction of one coordinate in a right-handed system, the new coordinate system becomes left-handed.

{\bf Adam.} Exactly, this introduces a minus sign in our definition of the curl (\ref{curlE}) which results in a minus sign in the dielectric properties (\ref{epmu}). So we need $\varepsilon=\mu=-1$. A material with such properties is negatively refracting. The idea of negative refraction with $\varepsilon=\mu=-1$ was introduced in 1968 \cite{ves68} where it was already clear that such materials exhibit some left-handedness and hence they were also called {\it left-handed}, but it took until 2006 when it was understood that they implement coordinate transformations from right-handed to left-handed systems \cite{GREE}. 

{\bf Gabriel.} How can we describe negative refraction in geometrical optics, which variables should I keep positive or negative?

{\bf Adam.} In electromagnetism you have $\varepsilon$ and $\mu$ to represent a material, in geometrical optics you can describe it simply by one quantity, the refractive index $n$, where $n^2=\varepsilon\mu$, so you have a choice of sign, plus or minus. For negative refraction you must use $n=-1$. Yet there is a problem with negative refraction.

{\bf Gabriel.} It is probably very hard to make.

{\bf Adam.} Negative refraction has been demonstrated in the optical range of the spectrum \cite{Zhang}, but perfect imaging with $\varepsilon=\mu=-1$ has never been achieved, only a ``poor-man's version'' of it \cite{poorman}. For a metal, one can have $\varepsilon=-1$ and for a metal sheet much thinner than the wavelength, one could ignore $\mu$ and so tolerate $\mu=1$. For such thin sheets the transfer of electromagnetic fields was demonstrated \cite{Fang}, but only over distances much shorter than the wavelength. 

{\bf Ben.} Is this the problem with negative refraction?

{\bf Adam.} There is a more fundamental problem. The device makes perfect copies of the electromagnetic field in three different regions of space, which is forbidden. 

{\bf Ben.} Does it violate the conservation of energy?

{\bf Adam.} Not necessarily in a situation where the object is continuously illuminated. 

{\bf Ben.} I know what the problem is: the field must jump instantaneously from one region to the other.

{\bf Adam.} Yes, it would violate causality. It cannot jump instantaneously, but it could settle over time to a regime where the field in all three regions is the same. On a short time scale the material cannot do what it does on a long time scale; the short-term response and the long-term response of the negatively-refracting material must be very different. This means that the material must be strongly frequency-dependent, {\it i.e.} strongly dispersive. If there is dispersion, there is also dissipation, which severely limits the imaging properties of negative-index materials. However, there are alternatives to perfect imaging with negative refraction, for example hyperlenses \cite{Narimanov} with an $\varepsilon$ tensor where one of the three eigenvalues is negative and the other two are positive. Another option is perfect imaging with positive refraction where the properties of curved geometries are used for imaging \cite{leo09}. 

{\bf Gabriel.} So far we discussed transformations of space, but we know from relativity that we should really regard space and time as combined in four-dimensional space-time. 

{\bf Adam.} Absolutely, let us discuss space-time. Our space-time coordinates are $x^\alpha$ where now $\alpha$ varies from 0 to 3, with $x_0=ct$ being the time coordinate. In space-time, the line element is
\begin{equation}
\text{d} s^2=\sum_{\alpha\beta}g_{\alpha\beta}\,\text{d} x^\alpha \text{d} x^\beta \,.
\end{equation}
For flat space we have the Minkowski metric
\begin{equation}
\text{d} s^2=c^2\text{d} t^2-\text{d} x^2-\text{d} y^2-\text{d} z^2 
\end{equation}
and thus the metric tensor is $g_{\alpha\beta}=\text{diag}(1,-1,-1,-1)$ in flat space. 

{\bf Ben.} What is the physical meaning of the space-time metric?

{\bf Adam.} It is a geometry of time. Let me explain: time is relative, so it depends on how one measures time. We define the proper time $\tau$ of a trajectory as the time measured by a clock moving on that trajectory. The important point is that the space-time line element $\text{d} s$ is independent of the frame of reference. In the co-moving frame $\text{d} x'^1 = \text{d} x'^2 = \text{d} x'^3 = 0$, and so $\text{d} s$ is proportional to the increment of time there. This means that the time read off from the moving clock, the proper time $\tau$, is the time given by the space-time metric:
\begin{equation}
\tau=\int\frac{\text{d} s}{c} \,.
\end{equation}
The geometry of space-time is the geometry of time.

{\bf Ben.} Is there a Fermat principle in space-time? 

{\bf Adam.} Yes, the proper time reaches an extremum for the actual trajectory of an object (usually a maximum, in contrast to Fermat's principle in three-dimensional space, as the space coordinates enter the metric with a minus sign). For light, this extremum of the proper time must also coincide with the requirement that $\text{d} s =0$, light lies on light cones in space-time.

{\bf Ben.} Can one understand a space-time geometry as a medium as well?

{\bf Adam.} Yes, one can show \cite{Plebanski} that the constitutive equations are
\begin{equation}
\bm{D}=\varepsilon_0\varepsilon\bm{E}+\frac{\bm{w}}{c}\times\bm{H},\quad \bm{B}=\mu_0\mu\bm{H}-\frac{\bm{w}}{c}\times\bm{E},
\end{equation}
where we have
\begin{equation}
w_i=\frac{g_{0i}}{g_{00}},\quad \varepsilon^{ij}=\mu^{ij}=\mp \frac{\sqrt{-g}}{g_{00}}g^{ij}.
\end{equation}

{\bf Gabriel.} I see that it is different from the materials we studied before.

{\bf Adam.} The materials that correspond to space-time geometries are magneto-electric materials, they mix electric and magnetic fields with a vector that depends on the off-diagonal elements of the space-time metric. The space-time geometry can be understood as an impedance-matched {\it moving} medium. Why do you think that this is a moving medium?

{\bf Gabriel.} Lorentz transformations from the rest frame to moving frames do the same, they mix magnetic and electric fields. 

{\bf Ben.} The moving medium responds to the electromagnetic field in its own rest frame

{\bf Gabriel.} --- so the field in the lab frame must be transformed to a co-moving frame. 

{\bf Ben.} I guess the velocity profile is $\bm{w}$.

{\bf Gabriel.} Yes, up to a prefactor.

\begin{figure}
\centering
\includegraphics[scale=0.6]{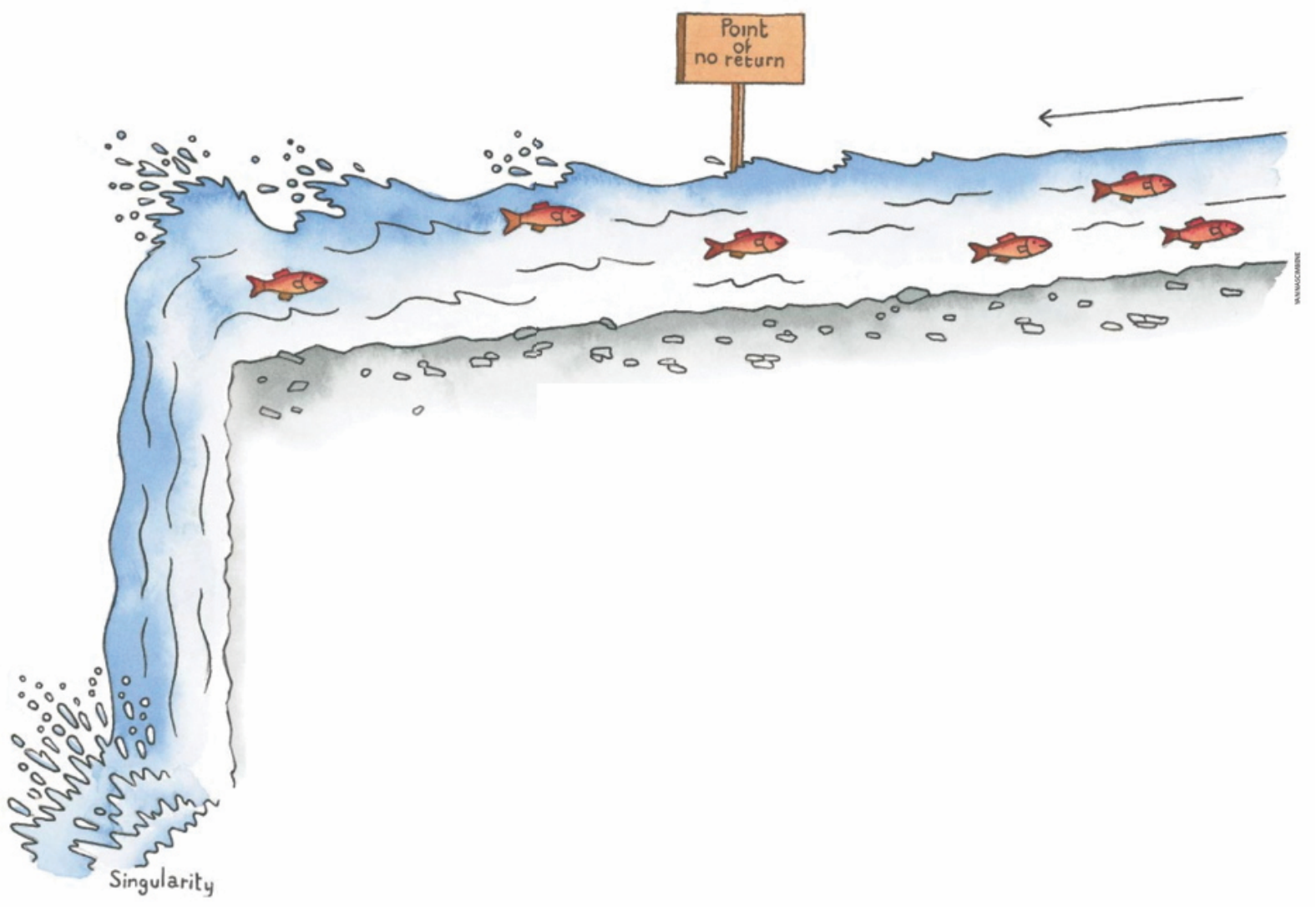}
\captionsetup{width=0.8\textwidth}
\caption{Aquatic analogue of the event horizon. (Reproduced with kindly permission of Yan Nascimbene.)}
\label{horizonfish}
\end{figure}

{\bf Adam.} Now I want to finish by giving you an example where a moving medium can be used to make something interesting: to build an analogue of the event horizon of a black hole. Black holes are perceived as sinister and mysterious, but they don't deserve this reputation. A black hole is actually something quite simple and its main mechanism operates not only in gravity, but also in many other situations in nature. Figure \ref{horizonfish} shows a very nice analogy of the black hole inspired by an idea of William Unruh \cite{SciAme}. Picture a river with flow velocity $u$ that gets faster and faster until the river reaches a waterfall: Singularity Falls. The water is populated with plenty of fish that, however, have one serious disadvantage: they have a maximal velocity, $c$. When the flow speed of the river is smaller than $c$ the fish can happily swim around, but in the region where the modulus of $u$ exceeds $c$ the fish can no longer swim upstream, they are swept along with the flow, drifting towards the inevitable: Singularity Falls. The point where the flow speed reaches $c$ is the event horizon.

{\bf Ben.} What does this example has to do with black holes?

{\bf Adam.} If we replace the fish by waves and work out their governing wave equation, it turns out to be the same as the wave equation in general relativity near a black hole horizon \cite{leo12}.

{\bf Gabriel.} How accurate is this correspondence?

\begin{figure}
\centering
\includegraphics[scale=0.6]{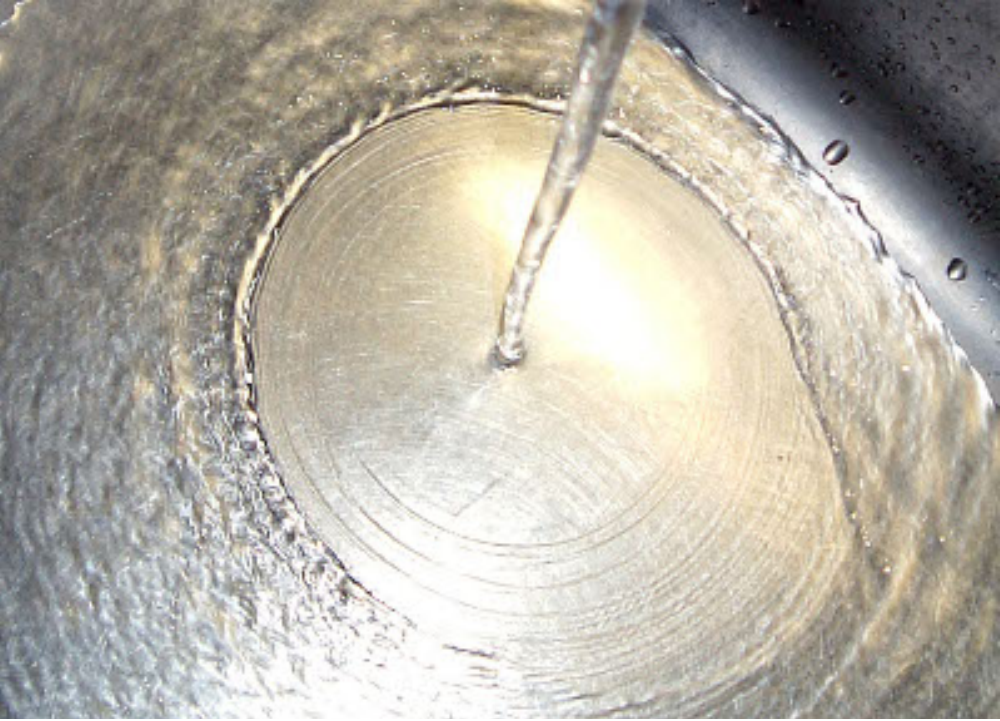}
\captionsetup{width=0.8\textwidth}
\caption{A water version of a white hole horizon caused by an {\it hydraulic jump} in a sink.}
\label{sink}
\end{figure}

{\bf Adam.} It is a mathematical equivalence, the wave propagation in the analogue and the original are described by the same equation, there is a one-to-one correspondence between them. You can even observe a closely related phenomenon in your kitchen sink. Suppose you open your tap, water is flowing out in a spout that hits a metal surface where it spreads out. If one looks at the water (Figure~\ref{sink}), one sees a ring structure around the point where the spout has hit the surface. Inside the ring the water surface is smooth, but outside of it waves appear. This is because inside of the ring the water flows faster than the speed of the waves, but outside of it the water has become sufficiently slow such that waves are formed. In fluid mechanics, this phenomenon is known as the {\it hydraulic jump}, in astrophysics it would be a {\it white-hole horizon}, which is the time reversal of the black-hole horizon.

{\bf Ben.} So, if we can make horizons with these simple experiments all the time, why would they be of scientific interest?

{\bf Adam.} The reason is quantum physics. There is a famous prediction by Stephen Hawking \cite{haw74}, where he claims that black holes are not truly black, but radiating. They turn vacuum fluctuations into real photons. According to quantum physics, the vacuum is not simply a desert of emptiness and nothing, but a sea of possibilities. The fluctuations of the quantum vacuum play a similar role credit plays in economy: they make things possible by borrowing. The horizon may borrow from the quantum vacuum repaying its debt with interest. This is the radiation the horizon emits. 

{\bf Gabrial.} I hear your words but I don't understand them. 

{\bf Ben.} Could we just test this in the laboratory? We replace the fish with light and the river with a moving dielectric medium. Hence, we could prove Hawking's hypothesis in the laboratory.

{\bf Gabriel.} But it seems almost impossible to make a medium move close to the speed of the waves, which in this case would be the speed of light. How can you do that?

{\bf Adam.} With a trick from nonlinear fiber optics \cite{Agrawal}. Consider an intense light pulse propagating through an optical fiber. The pulse induces a nonlinear response in the fiber glass that causes a change of the refractive index. This is known as the Kerr effect. The total refractive index in the fiber is given by:
\begin{equation}
n=n_0+n_2I\,,
\end{equation}
where $n_0$ is the linear refractive index, $n_2$ is the nonlinear coefficient and $I$ is the intensity of the light. Wherever the pulse is, it changes the refractive index, as if an additional piece of material is inserted there. This fictitious piece of material moves with the pulse and, as the pulse is a pulse of light, it moves with the speed of light. So it is absolutely elementary to make media moving at the speed of light (in the medium). It happens all the time in optical communication where light pulses travel through fibers. 

{\bf Gabriel.} But it is not exactly the same as in the picture of the fish swimming against the current. There you had a current getting faster in different regions of space, but the river was stationary, it does not change in time. Yet a light pulse is clearly non-stationary.

\begin{figure}
\centering
\includegraphics[scale=0.75]{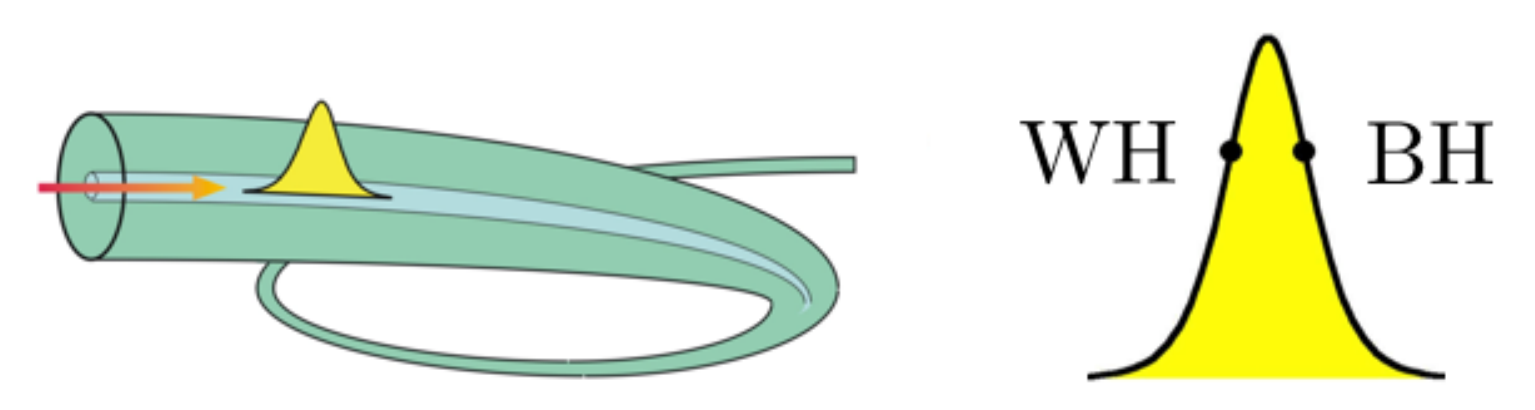}
\captionsetup{width=0.8\textwidth}
\caption{A light pulse in a fiber (left) changes the refractive index of the fiber and the {\it group velocity} of light waves inside it, this creates two horizons, corresponding to a white hole (WH) and a black hole (BH) at the leading and trailing parts of the pulse respectively.}
\label{fiberhorizon}
\end{figure}

{\bf Adam.} To see the connection to the picture of the river, imagine the fiber in the co-moving frame of the pulse. In this frame the light pulse is stationary and the fiber moves against the pulse with velocity $u$. The fiber is the river. Fast enough light waves, probes or ``fish'', propagate against the moving medium of the fiber with the initial velocity $c/n_0$. The light comes closer to the pulse where it experiences the Kerr effect; it propagates with the slower velocity $c/n$. If the refractive-index change induced by the pulse is strong enough to slow down the probe light such that $u=c/n$, than the probe experiences an horizon. As Figure ~\ref{fiberhorizon} illustrates, this change in the refractive index induces not one, but two horizons, as there are two regions where the condition $u=c/n$ would be fulfilled, one where the pulses rises and the other where the pulse falls.

{\bf Ben.} But why would the physics be different if we only think about the system in a different frame? I mean, it is still the same system.

{\bf Adam.} Well, exactly because the physics is the same in any reference frame we are allowed to think about the system in a frame of our choosing, like the co-moving one. It is just that sometimes it is easier to think about the physics in a particular frame. As our detectors are in the laboratory, we will need to transform back to the laboratory frame to see what the observable effects of the light pulse are. 

{\bf Gabriel.} Which physical quantities determine the creation of photons at the horizon?

{\bf Adam.} In astronomical black holes this quantity is the gravity gradient at the horizon. In optical black-hole analogues, the role of the gravity gradient is played by the refractive-index gradient due to the light pulse, $\partial_t \delta n$. One needs very short pulses, in the order of one cycle of light, to have a chance to see some interesting quantum effects of Hawking radiation in the laboratory. For optical pulses the cycle is about $1\, \mathrm{fs} = 10^{-15} \,\mathrm{s}$. Only for such short pulses we can possibly observe this effect.

{\bf Ben.} Is it feasible to achieve such short pulses?

{\bf Adam.} One can get $6\,\mathrm{fs}$ pulses from commercial ultrafast laser systems. When such pulses travel in a fiber they get further compressed, close to the single-cycle limit.

{\bf Gabriel.} How would you probe the horizon?

{\bf Adam.} The simplest way is to send in a continuous wave that chases after the pulse and look for some specific changes in its spectrum. When the probe wave is slowed down by the horizon to the extent that it is prevented from overtaking the pulse, it gets a bit squeezed behind the pulse, its wavelength gets shorter. One can see this on a spectrum analyser \cite{pkrs08}.

{\bf Gabriel.} But this is not Hawking radiation, is it?

{\bf Adam.} Well, a series of experiments \cite{bccg10,Negative} you can perform in order to prove several aspects of Hawking radiation. The ultimate one is the conceptually simplest one: just send nothing after the pulse in the fiber. The ``nothing'' is the quantum vacuum chasing after the pulse. Then look for the photons created from nothing. The photons should come in pairs, one from each side of the horizon. In astrophysics you have access to only one side of the horizon, but in laboratory analogues you can actually observe both photons. If you can measure the correlations of the two photons at the parts of the spectrum you expect you would demonstrate Hawking radiation in a way that is not only unachievable in astrophysics, but impossible. 

{\bf Gabriel.} What kind of spectrum you would expect from Hawking radiation?

{\bf Adam.} Actually in the ideal case we would get a thermal spectrum, just as for the astronomical black holes. In practice, this is not going to be the case.

{\bf Gabriel.} Could you show us something more concrete? I assume the mathematics of the matter could become rather complicated but I would like to see some proof.

{\bf Adam.} Fine, I can show you a simple case that includes most of the important features. Let us consider a one-dimensional moving medium...

{\bf Gabriel.} Is this a good approximation for a black hole? The black-hole horizon is a sphere, the Schwarzschild sphere. How can the Schwarzschild sphere be one-dimensional?

{\bf Adam.} Yes, it is an approximation, but for Hawking radiation there is only one relevant dimension, the radial direction. The reason is the following: near the horizon waves stand still, as their velocities are close to the speed of the moving medium. Yet these waves are oscillating, they are desperately trying to propagate. Their oscillations occur at shorter and shorter intervals the closer they are to the horizon. Even if the horizon is a curved surface, as the Schwarzschild sphere of astrophysical black holes, the wavelength is so short near the horizon that the other dimensions are completely negligible. Horizon physics is one-dimensional physics. 

{\bf Ben.} Understood.

{\bf Adam.} Let's assume that $\varepsilon=\mu=n$, {\it i.e.}\ impedance-matching. We know that impedance matching is essential for establishing an exact analogy between a medium and a geometry. 

{\bf Ben.} In a realistic fiber $\mu=1$, which is different from $\varepsilon$. Will this matter?

{\bf Adam.} Not much, because, in one dimension, violations of impedance matching just appear as reflections, as we have discussed before. Now, the Kerr effect modifies the refractive index, but not very much, say maximally $10^{-3}$. There is not going to be significant reflection. It is perfectly fine to assume impedance matching for making the calculations easier. Let me also assume that both the refractive index $n$ and the velocity $u$ do only depend on the distance $z$, but not on time $t$.

{\bf Gabriel.} But this is not true in the fiber either: here the pulse creates a time-depending refractive index.

{\bf Adam.} Sure, but not in the co-moving frame. There we agreed that the pulse stands still --- $n$ varies, but only in $z$ --- and the glass of the fiber moves towards us.

{\bf Ben.} So your theory will be valid for the fiber experiment in the co-moving frame of the pulse, but it may be much more general; it also describes other, hypothetical moving media. 

{\bf Adam.} Correct. Let me make one more simplification that is not necessary in principle, but makes the calculations even easier. Suppose that both the flow speed $|u|$ and the speed of light in the medium $c'$ is much smaller than $c$. As $c'$ is given by $c/n$ this means $n$ is much larger than unity. We cannot create such media with current technology, but the results for realistic materials will turn out to be the same. Let me go to the black board. Let's start from the action (\ref{action0}) with $\varepsilon=\mu=n(z)$ and assume that it is valid in a frame co-moving with the flow. 

{\bf Gabriel.} If $u$ varies there is no global co-moving frame in general. 

{\bf Adam.} You are right, but we can assign to each point of the medium a locally co-moving frame. The various frames do not necessarily join to form a global co-moving frame, but we can postulate that our Lagrangian for a medium at rest shall be valid in the locally co-moving frames.

{\bf Ben.} You simply assume the known laws of light propagation in each drop of a moving medium, in its own reference frame.

{\bf Adam.} Absolutely! Let me work out the mathematics. In the locally co-moving frame, the time derivative is $\partial_t + u\partial_z$, because changes in time may appear for two reasons there, a change in time $\partial_t$ in the laboratory frame and a change in time due to motion along a spatial variation, $u\partial_z$. So we take action (\ref{action0}) and simply replace $\partial_t A$ by $(\partial_t +u\,\partial_z )A$. In this way we get the action
\begin{equation}
S=\frac{\varepsilon_0}{2}\int\left( (\partial_t A+u\,\partial_z A)^2-c'^2(\partial_z A)^2\right)n\, \text{d} z\,\text{d} t\,.
\label{action1}
\end{equation}
Next we show that the field $A$ subject to action (\ref{action1}) experiences the moving medium as a geometry. Let us put $x^0=c't$ such that $\partial_t=c'\partial_0$. Let me write the quadratic form of the derivatives as $g^{\alpha\beta}(\partial_{\alpha} A)(\partial_\beta A)$ with the matrix
\begin{equation}
g^{\alpha\beta}=\frac{1}{c^2}\left(\begin{array}{cc}
   c'^2 & uc' \\
   uc' & u^2-c'^2 \\
  \end{array}\right).
\end{equation}
The notation is suggestive, $g^{\alpha\beta}$ is going to become the inverse metric tensor. We get for the determinant $g$ the expression
\begin{equation}
\frac{1}{g}=\mathrm{det}\left(g^{\alpha\beta}\right)=-\frac{c'^4}{c^4}=-\frac{1}{n^4}.
\end{equation}
For the time increment we get $n\text{d} t=n^2c^{-1}\text{d} x^0=\sqrt{-g}c^{-1}\text{d} x^0$ and so, finally, the action appears in geometric form
\begin{equation}
S=\frac{\varepsilon_0 c}{2}\int \sum_{\alpha\beta} g^{\alpha\beta}(\partial_{\alpha} A)(\partial_\beta A)\sqrt{-g}\,\text{d}^2 x\,.
\end{equation}
This equation for the action proves that the moving medium appears as a geometry. We also see this from the Euler-Lagrange equation for this action:
\begin{equation}
\sum_{\alpha\beta} \partial_\alpha\sqrt{-g}g^{\alpha\beta}\partial_\beta A=0\,.
\label{euler}
\end{equation}

{\bf Ben.} How do I see the geometry in this equation?

{\bf Adam.} Remember the expression we used for the divergence in curvilinear coordinates or curved geometries, Eq.~(\ref{divE}). You see that the left-hand side of the Euler-Lagrange equation (\ref{euler}) describes the divergence of a gradient, apart from the prefactor $1/\sqrt{-g}$ and the fact that we needed the minus sign in $g$, as $g$ is negative. The wave equation in moving media contains the d'Alembertian in a space-time geometry. Let us find out what this geometry is. We calculate the metric tensor by calculating the inverse of $g^{\alpha\beta}$...

{\bf Gabriel.} --- I will do it quickly. I get 
\begin{equation}
g_{\alpha\beta}=\frac{1}{c'^2}\left(\begin{array}{cc}
   c'^2-u^2 & uc' \\
   uc' & -c'^2 \\
  \end{array}\right)
\end{equation}

{\bf Adam.} --- and you see that the space-time metric is simply
\begin{equation}
 \text{d} s^2=c'^2\text{d} t^2-(\text{d} z-u\,\text{d} t)^2.
 \label{lightmetric}
\end{equation}

{\bf Gabriel.} That's a nice formula. It says that the space-time metric is the metric of light  $c'^2\text{d} t^2-\text{d} z'^2$ in the locally co-moving frames.

{\bf Adam.} Let me show you one more interesting aspect of one-dimensional moving media. We can define new coordinates $t_\pm$ with the increments
\begin{equation}
\text{d} t_\pm =\text{d} t -\frac{\text{d} z}{v_\pm}\quad\mbox{with}\quad v_\pm=\pm c'+u \,.
\label{tpm}
\end{equation}
Explicitly, the new coordinates are, in terms of the old ones
\begin{equation}
t_\pm(t,z) = t -\int\frac{\text{d} z}{v_\pm(z)} \,.
\end{equation}

{\bf Gabriel.} Obviously, $v_\pm$ is the added velocity of light and medium, with the plus sign for light propagating in positive direction and the minus sign for light in negative direction. 

{\bf Adam.} From the metric (\ref{lightmetric}) follows that $\text{d} t_+ \text{d} t_- = -v_+ v_- c'^2\text{d} s^2$. From this we read off the metric tensor in the $t_\pm$ coordinates:
\begin{equation}
g'_{\alpha\beta}=-\frac{v_+v_- c'^2}{2}\left(\begin{array}{cc}
	0  & 1 \\
   	1 & 0 \\
  \end{array}\right)\,,
\end{equation}
and get the inverse metric tensor
\begin{equation}
  g'^{\alpha\beta}=-\frac{2}{v_+v_- c'^2}\left(\begin{array}{cc}
	0  & 1 \\
   	1 & 0 \\
  \end{array}\right).
  \label{invtens}
\end{equation}
The determinant of $g'_{\alpha\beta}$ is given by $g'=-v_+^2 v_-^2 c'^4/4$. From this and Eq.~(\ref{invtens}) follows that $\partial_\alpha\sqrt{-g'}g'^{\alpha\beta}\partial_\beta = -\partial_+\partial_-$. As the wave equation (\ref{euler}) is invariant in any coordinate system we get 
\begin{equation}
\partial_+\partial_- A=0 \,,
\end{equation}
which has solutions of the form
\begin{equation}
A=A_+(t_+)+A_-(t_-)\,.
\end{equation}
This is d'Alembert's solution of wave propagation in free space; the wave can be a superposition of two wave packets, one as a function of $t_+$ and the other as a function of $t_-$. You see from definition (\ref{tpm}) that an amplitude $A$ at some $t_+$ value must move in $z$-space with velocity $v_+$: such amplitudes constitute a right-moving wave. Similarly, an amplitude at $t_-$ moves with $v_-$: it is left-moving. 

{\bf Gabriel.} Very nice.

{\bf Adam.} It gets even nicer. The wave does not only move like in free space, it moves in free space. Just define the coordinates $t'$ and $z'$ as
\begin{equation}
t'=\frac{t_++t_-}{2} \,,\quad z'=c\,\frac{t_--t_+}{2}
\end{equation}
such that
\begin{equation}
t_\pm=t'\mp\frac{z'}{c} \,.
\end{equation}
In these coordinates we get the free-space wave equation 
\begin{equation}
\left(\partial_{z'}^2-\frac{1}{c^2}\partial_{t'}^2\right) A=0 \,.
\end{equation}
The moving medium performs a coordinate transformation, it is a transformation medium. 

{\bf Ben.} We have seen something like this in the discussion of impedance matching. There we showed that a 1D impedance-matched medium is completely reflectionless, because it performs a coordinate transformation of space.

{\bf Adam.} Exactly! Now we have generalised this to moving media. Here a space-time transformation is necessary, a spatial transformation is not enough. The 1D moving medium is a space-time transformation medium. By the way, there even is a proposal \cite{McCall} followed by experiments on space-time cloaking where an event is hidden from observation. 

{\bf Gabriel.} But let's come back to horizons. I do not see how a horizon is special in the space-time transformation.

{\bf Adam.} The horizon cuts space-time into two parts, but otherwise the moving medium just performs a coordinate transformation. 

{\bf Gabriel.} But how can a coordinate transformation turn an empty wave, vacuum, into radiation? {\it Creatio ex nihilo?} How do you see this in your calculation? There is no Hawking radiation.

{\bf Ben.} I will check this in the lab.

{\bf Adam.} Hawking radiation comes from ...

{\bf Voice from above.} Adam, your time is up. 

\vspace{5mm}
\begin{center}
{\bf EPILOGUE}\\
{\it A glooming peace this morning with it brings;\\
The sun, for sorrow, will not show his head:\\
Go hence, to have more talk of these dark things -\\
That light appears where light has vanished:\\
For never was a story of such pedigree\\
Than this of light and of geometry.}
\end{center}
 
\section*{Acknowledgments}
This work was supported by the European Research Council, the Israeli Science Foundation and the Mexican Council of Science and Technology.

\end{document}